\documentclass[conference]{IEEEtran}
\IEEEoverridecommandlockouts
\usepackage{cite}
\usepackage{amsmath,amssymb,amsfonts}
\usepackage{algorithmic}
\usepackage{graphicx}
\usepackage{subcaption}
\usepackage{colortbl}
\usepackage{textcomp}
\usepackage{xcolor}
\usepackage{colortbl}
\usepackage[normalem]{ulem}
\usepackage{silence}
\usepackage{array}
\newlength\mylength
\def\BibTeX{{\rm B\kern-.05em{\sc i\kern-.025em b}\kern-.08em
    T\kern-.1667em\lower.7ex\hbox{E}\kern-.125emX}}

\usepackage{tabularx}

\begin{document}

\title{A Hybrid Deep Learning Architecture for Leukemic B-lymphoblast Classification}

\author{\IEEEauthorblockN{Sara Hosseinzadeh Kassani}
\IEEEauthorblockA{\textit{Department of Computer Science} \\
\textit{University of Saskatchewan}\\
Saskatoon, Canada \\
sara.kassani@usask.ca}
\and
\IEEEauthorblockN{Peyman Hosseinzadeh Kassani}
\IEEEauthorblockA{\textit{Department of Biomedical Engineering} \\
\textit{University of Tulane}\\
New Orleans, USA  \\
peymanhk@tulane.edu}
\and
\IEEEauthorblockN{Michal J. Wesolowski}
\IEEEauthorblockA{\textit{Department of Medical Imaging} \\
\textit{University of Saskatchewan}\\
Saskatoon, Canada \\
mike.wesolowski@usask.ca}
\and
\IEEEauthorblockN{Kevin A. Schneider}
\IEEEauthorblockA{\textit{Department of Computer Science} \\
	\textit{University of Saskatchewan}\\
	Saskatoon, Canada \\
	kevin.schneider@usask.ca}

\and
\IEEEauthorblockN{Ralph Deters}
\IEEEauthorblockA{\textit{Department of Computer Science} \\
	\textit{University of Saskatchewan}\\
	Saskatoon, Canada \\
	deters@cs.usask.ca}
}
\maketitle

\begin{abstract}
Automatic detection of leukemic B-lymphoblast cancer in microscopic images is very challenging due to the complicated nature of histopathological structures. To tackle this issue, an automatic and robust diagnostic system is required for early detection and treatment. In this paper, an automated deep learning-based method is proposed to distinguish between immature leukemic blasts and normal cells. The proposed deep learning based hybrid method, which is enriched by different data augmentation techniques, is able to extract high-level features from input images. Results demonstrate that the proposed model yields better prediction than individual models for Leukemic B-lymphoblast classification with 96.17\% overall accuracy, 95.17\% sensitivity and 98.58\% specificity. Fusing the features extracted from intermediate layers, our approach has the potential to improve the overall classification performance.
\end{abstract}

\begin{IEEEkeywords}
Acute Lymphoblastic Leukemia; Deep learning; Multi-model ensemble; Feature representation; Medical imaging
\end{IEEEkeywords}

\section{Introduction}
\begin{figure}[h]
 	\centering 
 	\begin{subfigure}{0.13\textwidth}
 		\includegraphics[width=\linewidth]{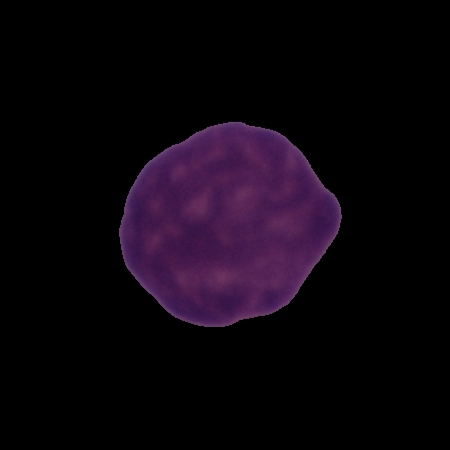}
 		\label{fig:1}
 	\end{subfigure}\hfil
 	\begin{subfigure}{0.13\textwidth}
 		\includegraphics[width=\linewidth]{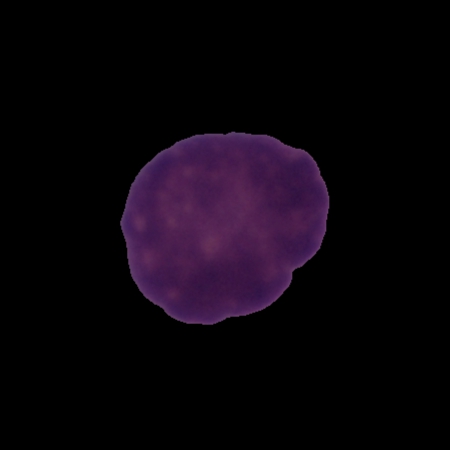}
 		\label{fig:2}
 	\end{subfigure}\hfil
 	\begin{subfigure}{0.13\textwidth}
 		\includegraphics[width=\linewidth]{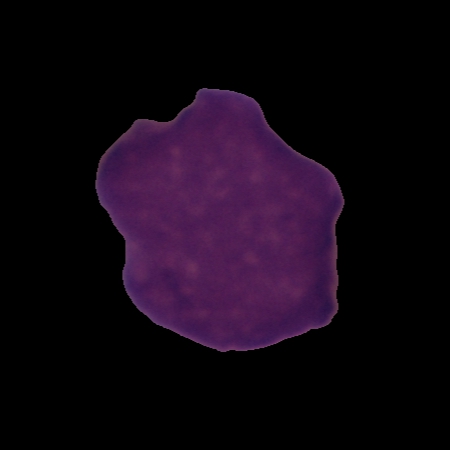}
 		\label{fig:3}
 	\end{subfigure}\hfil
 	\medskip
 	\begin{subfigure}{0.13\textwidth}
 		\includegraphics[width=\linewidth]{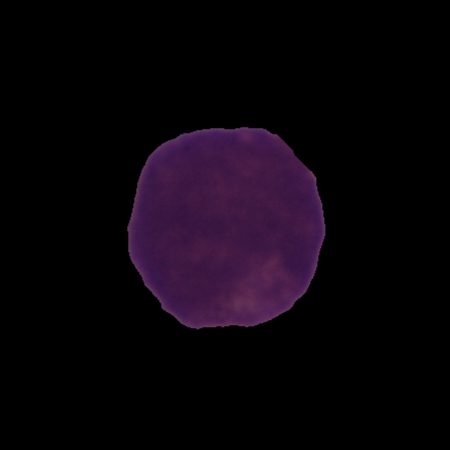}
 		\label{fig:4}
 	\end{subfigure}\hfil
 	\begin{subfigure}{0.13\textwidth}
 		\includegraphics[width=\linewidth]{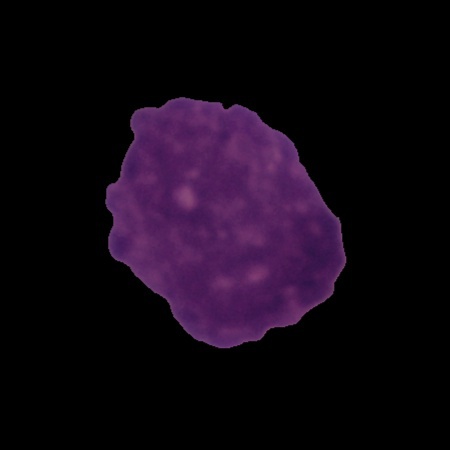}
 		\label{fig:5}
 	\end{subfigure}\hfil
 	\begin{subfigure}{0.13\textwidth}
 		\includegraphics[width=\linewidth]{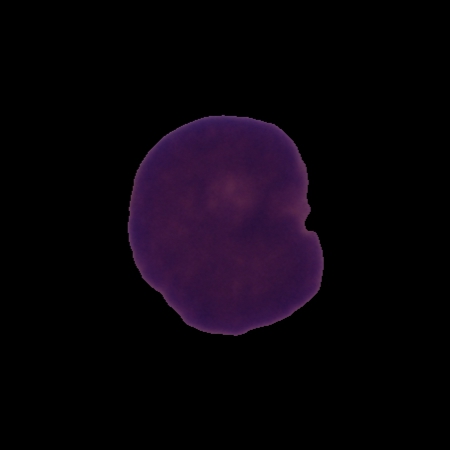}
 		\label{fig:6}
 	\end{subfigure}\hfil
 	
 	\caption{Normal B-lymphoid cells (top row), leukemic B-lymphoblast cells (bottom row). }
 	\label{fig:dataExamples}
 \end{figure}

Leukemia is a type of cancer associated with white blood cells that originates in the bone marrow and affects both children and adults. Leukemia can be divided into acute or chronic categories based on how quickly it progresses. There are four types of leukemia namely, Acute Myelogenous Leukemia (AML), Acute Lymphoblastic Leukemia (ALL), Chronic Myeloid Leukemia (CML), and Chronic Lymphocytic Leukemia (CLL)~\cite{LAOSAI2018127}~\cite{VOGADO2018415}. The most common types of leukemia that affect young children are AML and ALL. In ALL, lymphocytes - a type of white blood cell (WBC) - do not function properly and reproduce out of control, leading to anemia~\cite{tran2018blood}. This can lead to premature death if it is diagnosed in later stages or if the treatment process is delayed. Subject age is an important risk factor affecting prognosis, since the risk of developing ALL is highest in children below the age of 7-8 years. The risk then decreases until the mid-20s and begins to increase again after age 50. According to the data provided by~\cite{Leukemia-key-statistics-url}, in 2018, about 5930 new cases of ALL will be diagnosed and about 1500 patients are expected to die of ALL, including both children and adults, in the United States. The risk of getting ALL is slightly higher in males than females, and higher in whites than African-Americans. However, if leukemia is diagnosed in its early stages, it is highly curable and increases the survival rate of the patients. Considering the large-scale of histopathology images, assessment of the images in a conventional way can be laborious, error-prone and hugely time-consuming since some images are highly variable in morphology which is difficult to analyze. Therefore, developing accurate and reliable approaches for Leukemia detection is important for early treatment. Numerous study results showed that with the advancement of computational capabilities, hidden trends, patterns and relationships can be discovered using the application of data mining approaches in many different areas~\cite{DEABank, RADY2019100178,THOMAS2018160, Mardanisamani2019}. Fig~\ref{fig:dataExamples} illustrates examples of ALL and healthy cells.

\begin{figure}[t]
	\centering
	\includegraphics[width=0.3\textwidth]{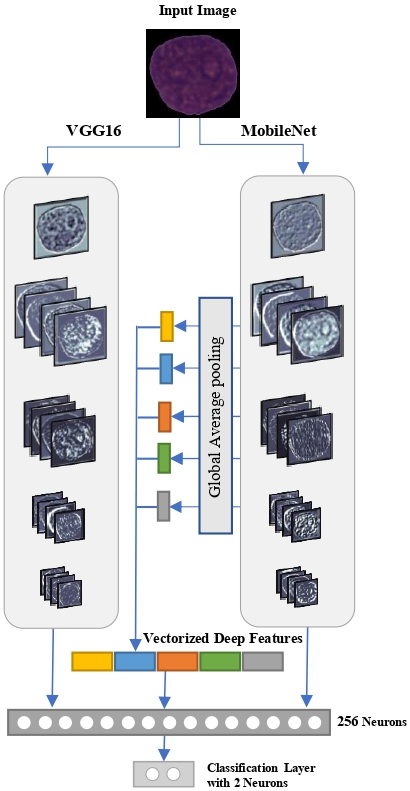}
	\caption{Architecture of the proposed CNN.}
	\label{fig:ProposedArchitecture}
\end{figure}

The details of our approach are shown in Fig~\ref{fig:ProposedArchitecture} and are described in the subsequent sections. Briefly, we present an automatic leukemic B-lymphoblast classification system using a hybrid of two Convolution Neural Network (CNN) and transfer learning to extract features from each input image. Unlike previous approaches, instead of using deep features extracted from the entire pre-trained architectures, in our approach, fusing the features from specific abstraction layers can be deemed as auxiliary features lead to further improvement of the classification accuracy. In this approach features extracted from the lower levels are combined into higher dimension feature maps to help improve the discriminative capability of intermediate features and also overcome the problem of network gradient vanishing/exploding.
\section{Related Studies}
Several methods for automated leukemia detection on microscopic images have been reported in the literature over the years. Singhal et al.~\cite{Singhal2014} have used a support vector machine (SVM) classifier for automatic detection of Acute Lymphoblastic Leukemia based on geometric features and local binary pattern (LBP) texture features. Experimental results showed that the LBP texture features perform better with 89.72\% accuracy compared to the shape features with 88.79\% one. 

The model proposed by Yu et al.~\cite{Yu2017} is a combination of state-of-the-art convolution neural networks including ResNet50, InceptionV3, VGG16, VGG19 and Xception for automatic cell recognition system using convolutional neural networks. The obtained result of the proposed model is compared to traditional machine learning algorithms such as K-Nearest Neighbors (KNN), Support Vector Machine (SVM), Logistic Regression (LR), and Decision Tree (DT). This approach resulted in 88.50\% accuracy with CNN models.

In the method proposed by Mohamed et al.~\cite{Mohamed2018}, the color space of each image is converted to YCbCr and then the gaussian distribution of Cb and Cr values is constructed. For training a classifier, various features such as texture, size and morphological are computed. The proposed model achieved 94.3\% accuracy using Random Forest classifier for diagnosing of Leukemia (ALL and AML) and Myeloma.

Mohapatra et al.~\cite{Mohapatra2014} described a method for acute leukemia detection in stained blood smear and bone marrow microscopic images. An ensemble model was trained using features extracted from the input images. The results showed that ensemble of classifiers achieved 94.73\% average prediction accuracy with an average sensitivity and average specificity of greater than 90\% in comparison with other standard classifiers, i.e., naive Bayesian (NB), K-nearest neighbor, multilayer perceptron (MLP), radial basis functional network (RBFN), and SVM.

In the method proposed by Patel et al.~\cite{PATEL2015}, leukemia detection was modeled by k-mean clustering. The model was also able to calculate the percentage of leukemia infection in microscopic images. The performance of Patel’s method was 93.57\% accuracy.

Finally, Mourya et al.~\cite{Mourya2018} proposed the use of a deep learning-based hybrid architecture, with two CNN architectures to improve the classification accuracy. The model was tested on 636 samples of normal and ALL cells and showed 89.70\% accuracy. 
\section{Materials and Methods}
\subsection{Methodology}
Our approach consists of the following stages: Initially, we enhance the quality of visual information of each input image using different pre-processing and augmentation techniques to increase the visibility of crucial structures. Once input images are prepared, they are used in the feature extraction phase with the proposed hybrid architecture. We explore two architectures namely, VGG16~\cite{VGGNet} and MobileNet~\cite{MobileNet2017} for our hybrid model. VGG16 is a very simple yet effective architecture consists of 13 convolutional using 3 x 3 convolution filters followed by max pooling layers and two 4096 fully-connected layers, followed by a softmax classifier. MobileNet architecture is designed for object recognition on mobile devices. This architecture consists of depth-wise separable convolution and 1×1 point-wise convolutions. The performance of the MobileNet architecture is evaluated on ImageNet dataset and achieved an accuracy in the same level of accuracy as VGG16 with 32 times less parameters while is 27 times less computationally intensive. Since each architecture has its own shortcomings, we come up with an integrating strategy to make use of the advantages of both architectures in order to improve overall prediction accuracy. The extracted features were trained by a multi-layer perceptron to classify each image into corresponding class probabilities. Finally, the performance of the proposed architecture is evaluated on test images.

\subsection{Experimental Dataset}\label{sec:ExperimentalDataset}
The dataset used for this study is based on Classification of Normal versus Malignant Cells in B-ALL White Blood Cancer Microscopic Images as part of ISBI 2019 challenge provided by SBILab which is available for the public at~\cite{SBILab-url}. The images are stored with the resolution of 450×450 pixels using the 24-bit RGB color system. The size of each cell is approximately the size of 300×300 pixels. The images were annotated by experienced oncologist for the classification procedure. The methods developed by~\cite{34-Anubha, 35-Gupta2017, 36-Duggal, 37-Duggal2017, 38-Rahul2016} is employed for segmentation and stain normalization of the provided dataset. The dataset contains a total of 76 individual subjects (47 ALL subjects and 29 Normal subjects), containing a total cells images of 7272 ALL and 3389 normal cells. 

\subsection{Data Pre-processing}\label{sec:preProcessing}
\subsubsection{Normalization}
Two normalization methods are used in this experiment to compare the performance of different methods. First, we subtract the mean RGB value of all images from the training set divided by its standard deviation to normalize the input images as suggested in~\cite{21-Yu2017}. We also normalize images using ImageNet mean subtraction as a pre-processing step. The ImageNet mean is a pre-computed constant derived from ImageNet database~\cite{Krizhevsky2012}.

\subsubsection{Resizing}
Regarding to the black margin of each image as illustrated in Fig\ref{fig:dataExamples}, we resized all images from the image center of the original size of 450×450 pixels to the appropriate size 380×380 pixels using bicubic interpolation to ensure each cell is located at the center and reduce the non-informative adjacent background regions.

\subsubsection{Data Augmentation}\label{sec:DataAugmentation}
CNNs demonstrated state-of-the-art performance in different tasks~\cite{CHOI2019,OKTAY2019,LI2019}. However, the performance of CNNs highly depends on training data size. Due to the data privacy issues in medical domain, collecting adequate clinical images is a challenge. To address the issue of limited dataset size and avoid over-fitting problems, we applied various data augmentation techniques to optimize the CNN performance as suggested in recent studies~\cite{kassaniaug2019}~\cite{pham2018deep} including contrast adjustments and brightness correction, horizontal and vertical flips and intensity adjustments. The class distributions of dataset before and after data augmentation is presented in Table 1. 
\begin{table}[hb]
	\centering
	\caption{Total number of class distributions before and after data augmentation.}\label{tab:DatasetDistribution}
\def\arraystretch{1.3}
\begin{tabular}{|l|c|c|}
	\hline
	\multicolumn{3}{|c|}{Number of images}                                                             \\ \hline
	Cell type     & \multicolumn{1}{l|}{Before augmentation} & \multicolumn{1}{l|}{After augmentation} \\ \hline
	Healthy cells & 3389                                     & 27930                                   \\ \hline
	ALL cells     & 7272                                     & 53591                                   \\ \hline
\end{tabular}
\end{table}

\subsection{Proposed Deep CNN Architecture with Auxiliary Components}
The main contribution of our approach is proposing a hybrid CNN model that combines low-level features from intermediate layers in order to generate high-level discriminative feature maps for immature leukemic blast classification. In this approach, two well-established CNN architectures, namely MobileNet and VGG16 which have shown excellent performance in many computer vision tasks are used~\cite{SANTIN2019, CHEVTCHENKO}. For VGG16 architecture, the initial weights are obtained from weights learned from ImageNet by transfer learning strategy. As illustrated in Figure 2, from MobileNet architecture, features from five convolution layers are extracted. Then each of them followed by an average pooling layer. Next, we concatenated them into a single feature vector. Thereafter, we connect a new fully connected (FC) layer with 256 hidden units with rectified linear unit (ReLU) activation function. Finally, two output neurons associating with normal and malignant cases with softmax non-linearity activation function are used at the classifier layer. These extracted features from selected intermediate layers can act as a complementary set of features to learn highly discriminative features beside the existing extracted deep features. This approach results in detection of more complex patterns from each input image and gives higher accuracy with lower error rate. Employing very deep architecture for training limited samples could have the issues of vanishing gradients and poor local minima. The main benefit of applying a global average pooling layer is reducing the number of parameters in very deep architectures. This reduction helps to prevent getting stuck in the poor local minima in a high dimensional space which often occurs in learning from very deep CNNs. Additionally, once the number of parameters decreases, we can ensure the gradient flow within the deep network and hence the learning process becomes stable regardless of the network depth, i.e. the number of hidden layers.
\subsection{Evaluation Metrics}
To evaluate the performance of the proposed method, three mostly used evaluation metrics namely, accuracy, sensitivity and specificity are considered. Accuracy shows the number of correctly classified ALL cases divided by the total number of test images denoting the overall correctness, is defined as:
\begin{equation}
Accuracy (\%)= \frac{TP+TN}{TP+TN+FP+FN} \times 100
\end{equation}

In detecting disease, sensitivity or True Positive Rate (TPR) is a measure of the proportion of true positive results to all real positives (subjects that have the disease). If cancer samples in the provided dataset are limited, the model has to be sensitive.
\begin{equation}
Sensitivity(\%)= \frac{TP}{TP+FN} \times 100
\end{equation}
Specificity or True Negative Rate (TNR) is a measure of the true proportion of negative results to all real negatives (subjects that do not have the disease). High specificity means that the model is good in detecting healthy cases.

\begin{equation}
Specificity (\%)= \frac{TN}{TN+FP} \times 100
\end{equation}

\section{Experiment and Results}
\subsection{Experimental Setup}
For our experiments, 70\% of the images of each class are assigned to the training set, 20\% to the validation set, and the remaining 10\% to the test set. To obtain the optimal accuracy, several hyper-parameter tuning, using an exhaustive grid-search, is utilized. The effect of different optimizers, namely adaptive moment estimation (Adam), stochastic gradient descent (SGD) with momentum, and root mean square propagation (RMSProp) are investigated. For SGD optimizer, the momentum term was set to 0.9. For Adam optimizer, $\beta 1$ and $\beta 2$ were set to 0.7 and 0.999, respectively. For RMSProp optimizer, rho and  $ \epsilon $ were set to 0.8 and None, respectively. The learning rate was set to 0.001 for the Adam optimizer and to 0.0001 for both RMSProp and SGD optimizer. We utilized ReLU activation function and dropout~\cite{srivastava2014dropout} in the fully-connected layer with a rate of 0.4 to prevent over-fitting. The batch size was set to 32 in order to fit into the GPU memory. All models are trained for 1000 epochs. Our experiment is implemented in Python using the Keras package with Tensorflow as the deep learning framework backend and run on Nvidia GeForce GTX 1080 Ti GPU with 11GB RAM.

\subsection{Results and Discussion}
The obtained results are derived from the 967 test images of the ISBI 2019 challenge which were not used in the training phase. These test set are consist of 312 normal cases and 655 ALL cases. We first examine the effect of image normalization and different optimizers on the classification performance. The accuracy, sensitivity and specificity of the obtained results are tabulated in Table~\ref{tab:OptimizersNormalization}. 
\begin{table}[hb]
	\centering
	\caption{Effects of different optimizers and image normalization techniques. Bold values indicate the best result.}\label{tab:OptimizersNormalization}
	\def\arraystretch{1.3}
	\begin{tabular}{|l|l|l|l|l|}
		\hline
		Optimizer & Normalization & Accuracy & Sensitivity & Specificity \\ \hline
		Adam      & ImageNet Mean & 95.14    & \textbf{95.92}       & 93.44       \\ \hline
		RMSProp   & ImageNet Mean & 93.38    & 94.17       & 91.61       \\ \hline
		SGD       & ImageNet Mean & 93.17    & 91.3        & 98.42       \\ \hline
		Adam      & Dataset Mean  & \textbf{96.17}    & 95.17       & 98.58       \\ \hline
		RMSProp   & Dataset Mean  & 92.04    & 90.58       & 96.07       \\ \hline
		SGD       & Dataset Mean  & 89.76    & 86.96       & \textbf{99.53}        \\ \hline
	\end{tabular}
\end{table}

As the results confirm, there is a level of variation in all results when running the experiments with different optimizers and image normalization techniques. Analyzing Table~\ref{tab:OptimizersNormalization}, we observe that proposed model delivered high accuracy (96.17\%) on dataset mean normalization with Adam optimizer. High sensitivity (95.92\%) result achieved by Adam optimizer, and ImageNet mean normalization method, and high specificity (99.53\%) obtained by dataset mean normalization and SGD optimizer. Surprisingly, the worst classifier is observed by SGD optimizer and dataset mean normalization with an accuracy of 89.76\%, sensitivity of 86.96\%, and specificity of 99.53 (the last row in Table~\ref{tab:OptimizersNormalization}).

To justify the performance of the proposed approach, the performance of each architecture is individually evaluated. Table~\ref{tab:FinalResults} provides the comparison of the individual VGG16 and MobileNet architectures with the proposed model. From Table~\ref{tab:FinalResults}, it can be seen that our proposed method significantly outperforms the individual architectures on the provided dataset. Our model improves VGG16 up to 16\% and MobileNet by 8.17\% in terms of accuracy, which is considered significant. Moreover, the plain MobileNet architecture (88.00\%) gives a better performance than VGG16 architecture (80.77\%). This means the gap in accuracy is 7.23\%, in favor of MobileNet. This is probably because of the benefit of the depth-wise and point-wise blocks in MobileNet compared to regular convolutional blocks in VGG16. 
\begin{table}[hb]
	\centering
	\caption{Classification results from plain pre-trained networks and proposed model. Bold values indicate the best result.}\label{tab:FinalResults}
	\def\arraystretch{1.3}
	\begin{tabular}{|l|l|l|l|}
		\hline
		& Accuracy & Sensitivity & Specificity \\ \hline
		Plain MobileNet & 88       & 86.66       & 92.24       \\ \hline
		Plain VGG16     & 80.77    & 78.21       & 96.32       \\ \hline
		\textbf{Proposed model}  & \textbf{96.17}    & \textbf{95.17}       & \textbf{98.58}       \\ \hline
	\end{tabular}
\end{table}

For the sake of comparison, our proposed ensemble is compared with some of the recent studies in the literature in Table~\ref{tab:ComparativeResults}. As shown in Table~\ref{tab:ComparativeResults}, the proposed approach achieves better performance compared to other studies in terms of the accuracy. 

\begin{table}[ht]
	\centering
	\caption{The comparison of the proposed method with recent studies on leukemia detection}\label{tab:ComparativeResults}
	\def\arraystretch{1.3}
\begin{tabular}{|l|l|l|l|}
	\hline
	Dataset  & Method          & Accuracy & Year \\ \hline
	DTH      & Yu et al.~\cite{Yu2017}        & 88.50\%  & 2017 \\ \hline
	ISBI     & Mourya et al.~\cite{Mourya2018}          & 89.62\%  & 2018 \\ \hline
	ALL-IDB2 & Singhal et al.~\cite{Singhal2014}    & 89.72\%  & 2014 \\ \hline
	MISP     & Mohamed et al.~\cite{Mohamed2018}   & 93.00\%  & 2018 \\ \hline
	ALL-IDB  & Patel et al.~\cite{PATEL2015}     & 93.75\%  & 2015 \\ \hline
	IGH      & Mohapatra et al.~\cite{Mohapatra2014} & 94.73\%  & 2013 \\ \hline
	ISBI     & Proposed Approach          & \textbf{96.17}   & 2019 \\ \hline
\end{tabular}
\end{table}

The experimental results in Table~\ref{tab:ComparativeResults} confirm that the proposed ensemble, by aggregating features from intermediate layers outperforms all counterparts and achieves the highest accuracy. This indicates the important role of ensemble based deep learning in joint with highly descriptive feature. Our proposed learner gains accuracy of 96.17\% on the recent ISBI 2019 dataset while counterpart study at~\cite{Mourya2018} from the Table 4, gains accuracy of 89.62\% on the same dataset.

\section{Conclusion}
We presented an automatic CNN hybrid method for classification of ALL and healthy cells. Two well-established CNN, namely, VGG16 and MobileNet are used to extract features from multiple abstraction levels. Fusing the features from selected intermediate layers can be regarded as an auxiliary set of features which leads to further improvement of the classification accuracy. This approach not only helps to learn more complex patterns but also addresses the issues of vanishing gradients and poor local minima by reducing the number of parameters. The obtained results suggest that combining features learned by deep models improves the performance and yield more accurate result (96.17\%) than individual state-of-the-art networks. For future research directions, we intend to employ the ensemble of other CNN architectures to observe the change in accuracy.

\bibliographystyle{IEEEtran}
\bibliography{ref}

\begin{thebibliography}{10}
\providecommand{\url}[1]{#1}
\csname url@samestyle\endcsname
\providecommand{\newblock}{\relax}
\providecommand{\bibinfo}[2]{#2}
\providecommand{\BIBentrySTDinterwordspacing}{\spaceskip=0pt\relax}
\providecommand{\BIBentryALTinterwordstretchfactor}{4}
\providecommand{\BIBentryALTinterwordspacing}{\spaceskip=\fontdimen2\font plus
\BIBentryALTinterwordstretchfactor\fontdimen3\font minus
  \fontdimen4\font\relax}
\providecommand{\BIBforeignlanguage}[2]{{%
\expandafter\ifx\csname l@#1\endcsname\relax
\typeout{** WARNING: IEEEtran.bst: No hyphenation pattern has been}%
\typeout{** loaded for the language `#1'. Using the pattern for}%
\typeout{** the default language instead.}%
\else
\language=\csname l@#1\endcsname
\fi
#2}}
\providecommand{\BIBdecl}{\relax}
\BIBdecl

\bibitem{LAOSAI2018127}
\BIBentryALTinterwordspacing
J.~Laosai and K.~Chamnongthai, ``Classification of acute leukemia using
  medical-knowledge-based morphology and cd marker,'' \emph{Biomedical Signal
  Processing and Control}, vol.~44, pp. 127 -- 137, 2018. [Online]. Available:
  \url{http://www.sciencedirect.com/science/article/pii/S1746809418300272}
\BIBentrySTDinterwordspacing

\bibitem{VOGADO2018415}
\BIBentryALTinterwordspacing
L.~H. Vogado, R.~M. Veras, F.~H. Araujo, R.~R. Silva, and K.~R. Aires,
  ``Leukemia diagnosis in blood slides using transfer learning in cnns and svm
  for classification,'' \emph{Engineering Applications of Artificial
  Intelligence}, vol.~72, pp. 415 -- 422, 2018. [Online]. Available:
  \url{http://www.sciencedirect.com/science/article/pii/S0952197618301039}
\BIBentrySTDinterwordspacing

\bibitem{tran2018blood}
T.~Tran, O.-H. Kwon, K.-R. Kwon, S.-H. Lee, and K.-W. Kang, ``Blood cell images
  segmentation using deep learning semantic segmentation,'' in \emph{2018 IEEE
  International Conference on Electronics and Communication Engineering
  (ICECE)}.\hskip 1em plus 0.5em minus 0.4em\relax IEEE, 2018, pp. 13--16.

\bibitem{Leukemia-key-statistics-url}
\BIBentryALTinterwordspacing
``{Key Statistics for Acute Lymphocytic Leukemia (ALL)}.'' [Online]. Available:
  \url{https://www.cancer.org/cancer/acute-lymphocytic-leukemia/about/key-statistics.html}
\BIBentrySTDinterwordspacing

\bibitem{DEABank}
S.~E.~N. {Sara Hosseinzadeh Kassani, Peyman Hosseinzadeh Kassani},
  ``{Introducing a hybrid model of DEA and data mining in evaluating
  efficiency. Case study: Bank Branches},'' \emph{Academic Journal of Research
  In Economics and Management}, vol.~3, no.~2, 2015.

\bibitem{RADY2019100178}
\BIBentryALTinterwordspacing
E.-H.~A. Rady and A.~S. Anwar, ``Prediction of kidney disease stages using data
  mining algorithms,'' \emph{Informatics in Medicine Unlocked}, vol.~15, p.
  100178, 2019. [Online]. Available:
  \url{http://www.sciencedirect.com/science/article/pii/S2352914818302387}
\BIBentrySTDinterwordspacing

\bibitem{THOMAS2018160}
\BIBentryALTinterwordspacing
M.~C. Thomas, W.~Zhu, and J.~A. Romagnoli, ``Data mining and clustering in
  chemical process databases for monitoring and knowledge discovery,''
  \emph{Journal of Process Control}, vol.~67, pp. 160 -- 175, 2018, big Data:
  Data Science for Process Control and Operations. [Online]. Available:
  \url{http://www.sciencedirect.com/science/article/pii/S095915241730032X}
\BIBentrySTDinterwordspacing

\bibitem{Mardanisamani2019}
\BIBentryALTinterwordspacing
S.~Mardanisamani, F.~Maleki, S.~H. Kassani, S.~Rajapaksa, H.~Duddu, M.~Wang,
  S.~Shirtliffe, S.~Ryu, A.~Josuttes, T.~Zhang, S.~Vail, C.~Pozniak, I.~Parkin,
  I.~Stavness, and M.~Eramian, ``{Crop Lodging Prediction from UAV-Acquired
  Images of Wheat and Canola using a DCNN Augmented with Handcrafted Texture
  Features},'' jun 2019. [Online]. Available:
  \url{http://arxiv.org/abs/1906.07771}
\BIBentrySTDinterwordspacing

\bibitem{Singhal2014}
V.~Singhal and P.~Singh, ``{Local Binary Pattern for automatic detection of
  Acute Lymphoblastic Leukemia},'' in \emph{2014 20th National Conference on
  Communications, NCC 2014}, 2014.

\bibitem{Yu2017}
\BIBentryALTinterwordspacing
W.~Yu, J.~Chang, C.~Yang, L.~Zhang, H.~Shen, Y.~Xia, and J.~Sha, ``{Automatic
  classification of leukocytes using deep neural network},'' in \emph{2017 IEEE
  12th International Conference on ASIC (ASICON)}.\hskip 1em plus 0.5em minus
  0.4em\relax IEEE, oct 2017, pp. 1041--1044. [Online]. Available:
  \url{http://ieeexplore.ieee.org/document/8252657/}
\BIBentrySTDinterwordspacing

\bibitem{Mohamed2018}
\BIBentryALTinterwordspacing
H.~Mohamed, R.~Omar, N.~Saeed, A.~Essam, N.~Ayman, T.~Mohiy, and A.~AbdelRaouf,
  ``{Automated detection of white blood cells cancer diseases},'' in \emph{2018
  First International Workshop on Deep and Representation Learning
  (IWDRL)}.\hskip 1em plus 0.5em minus 0.4em\relax IEEE, mar 2018, pp. 48--54.
  [Online]. Available: \url{https://ieeexplore.ieee.org/document/8358214/}
\BIBentrySTDinterwordspacing

\bibitem{Mohapatra2014}
S.~Mohapatra, D.~Patra, and S.~Satpathy, ``{An ensemble classifier system for
  early diagnosis of acute lymphoblastic leukemia in blood microscopic
  images},'' \emph{Neural Computing and Applications}, 2014.

\bibitem{PATEL2015}
\BIBentryALTinterwordspacing
N.~Patel and A.~Mishra, ``Automated leukaemia detection using microscopic
  images,'' \emph{Procedia Computer Science}, vol.~58, pp. 635 -- 642, 2015,
  second International Symposium on Computer Vision and the Internet
  (VisionNet’15). [Online]. Available:
  \url{http://www.sciencedirect.com/science/article/pii/S1877050915021936}
\BIBentrySTDinterwordspacing

\bibitem{Mourya2018}
\BIBentryALTinterwordspacing
S.~Mourya, S.~Kant, P.~Kumar, A.~Gupta, and R.~Gupta, ``{LeukoNet: DCT-based
  CNN architecture for the classification of normal versus Leukemic blasts in
  B-ALL Cancer},'' oct 2018. [Online]. Available:
  \url{http://arxiv.org/abs/1810.07961}
\BIBentrySTDinterwordspacing

\bibitem{VGGNet}
\BIBentryALTinterwordspacing
K.~Simonyan and A.~Zisserman, ``{Very Deep Convolutional Networks for
  Large-Scale Image Recognition},'' \emph{CoRR}, vol. abs/1409.1, 2014.
  [Online]. Available: \url{http://arxiv.org/abs/1409.1556}
\BIBentrySTDinterwordspacing

\bibitem{MobileNet2017}
\BIBentryALTinterwordspacing
A.~G. Howard, M.~Zhu, B.~Chen, D.~Kalenichenko, W.~Wang, T.~Weyand,
  M.~Andreetto, and H.~Adam, ``{MobileNets: Efficient Convolutional Neural
  Networks for Mobile Vision Applications},'' apr 2017. [Online]. Available:
  \url{http://arxiv.org/abs/1704.04861}
\BIBentrySTDinterwordspacing

\bibitem{SBILab-url}
\BIBentryALTinterwordspacing
``{Classification of Normal vs Malignant Cells in B-ALL White Blood Cancer
  Microscopic Images:ISBI 2019}.'' [Online]. Available:
  \url{https://competitions.codalab.org/competitions/20429}
\BIBentrySTDinterwordspacing

\bibitem{34-Anubha}
G.~Anubha, D.~Rahul, G.~Ritu, K.~Lalit, T.~Nisarg, and S.~Devprakash,
  ``{GCTI-SN: Geometry-Inspired Chemical and Tissue Invariant Stain
  Normalization of Microscopic Medical Images}.''

\bibitem{35-Gupta2017}
\BIBentryALTinterwordspacing
R.~Gupta, P.~Mallick, R.~Duggal, A.~Gupta, and O.~Sharma, ``{Stain Color
  Normalization and Segmentation of Plasma Cells in Microscopic Images as a
  Prelude to Development of Computer Assisted Automated Disease Diagnostic Tool
  in Multiple Myeloma},'' \emph{Clinical Lymphoma Myeloma and Leukemia},
  vol.~17, no.~1, p. e99, feb 2017. [Online]. Available:
  \url{https://linkinghub.elsevier.com/retrieve/pii/S2152265017304688}
\BIBentrySTDinterwordspacing

\bibitem{36-Duggal}
\BIBentryALTinterwordspacing
R.~Duggal, A.~Gupta, R.~Gupta, M.~Wadhwa, and C.~Ahuja, ``{Overlapping Cell
  Nuclei Segmentation in Microscopic Images Using Deep Belief Networks},'' in
  \emph{Proceedings of the Tenth Indian Conference on Computer Vision, Graphics
  and Image Processing}, ser. ICVGIP '16.\hskip 1em plus 0.5em minus
  0.4em\relax New York, NY, USA: ACM, 2016, pp. 82:1----82:8. [Online].
  Available: \url{http://doi.acm.org/10.1145/3009977.3010043}
\BIBentrySTDinterwordspacing

\bibitem{37-Duggal2017}
R.~Duggal, A.~Gupta, R.~Gupta, and P.~Mallick, ``{SD-Layer: Stain
  Deconvolutional Layer for CNNs in Medical Microscopic Imaging BT - Medical
  Image Computing and Computer-Assisted Intervention − MICCAI 2017},''
  M.~Descoteaux, L.~Maier-Hein, A.~Franz, P.~Jannin, D.~L. Collins, and
  S.~Duchesne, Eds.\hskip 1em plus 0.5em minus 0.4em\relax Cham: Springer
  International Publishing, 2017, pp. 435--443.

\bibitem{38-Rahul2016}
D.~Rahul, G.~Anubha, and R.~Gupta, ``{Segmentation of overlapping/touching
  white blood cell nuclei using artificial neural networks},'' \emph{CME Series
  on Hemato-Oncopathology, All India Institute of Medical Sciences (AIIMS)},
  2016.

\bibitem{21-Yu2017}
Z.~Yu, X.~Jiang, T.~Wang, and B.~Lei, ``{Aggregating Deep Convolutional
  Features for Melanoma Recognition in Dermoscopy Images BT - Machine Learning
  in Medical Imaging},'' Q.~Wang, Y.~Shi, H.-I. Suk, and K.~Suzuki, Eds.\hskip
  1em plus 0.5em minus 0.4em\relax Cham: Springer International Publishing,
  2017, pp. 238--246.

\bibitem{Krizhevsky2012}
A.~Krizhevsky and G.~E. Hinton, ``{ImageNet Classification with Deep
  Convolutional Neural Networks},'' \emph{Neural Information Processing
  Systems}, 2012.

\bibitem{CHOI2019}
\BIBentryALTinterwordspacing
J.-H. Choi and J.-S. Lee, ``Embracenet: A robust deep learning architecture for
  multimodal classification,'' \emph{Information Fusion}, vol.~51, pp. 259 --
  270, 2019. [Online]. Available:
  \url{http://www.sciencedirect.com/science/article/pii/S1566253517308242}
\BIBentrySTDinterwordspacing

\bibitem{OKTAY2019}
\BIBentryALTinterwordspacing
A.~B. Oktay and A.~Gurses, ``Automatic detection, localization and segmentation
  of nano-particles with deep learning in microscopy images,'' \emph{Micron},
  vol. 120, pp. 113 -- 119, 2019. [Online]. Available:
  \url{http://www.sciencedirect.com/science/article/pii/S0968432818304013}
\BIBentrySTDinterwordspacing

\bibitem{LI2019}
\BIBentryALTinterwordspacing
L.~Li, X.~Zhao, W.~Lu, and S.~Tan, ``Deep learning for variational
  multimodality tumor segmentation in pet/ct,'' \emph{Neurocomputing}, 2019.
  [Online]. Available:
  \url{http://www.sciencedirect.com/science/article/pii/S0925231219304667}
\BIBentrySTDinterwordspacing

\bibitem{kassaniaug2019}
\BIBentryALTinterwordspacing
S.~H. Kassani and P.~H. Kassani, ``A comparative study of deep learning
  architectures on melanoma detection,'' \emph{Tissue and Cell}, vol.~58, pp.
  76 -- 83, 2019. [Online]. Available:
  \url{http://www.sciencedirect.com/science/article/pii/S0040816619300904}
\BIBentrySTDinterwordspacing

\bibitem{pham2018deep}
T.-C. Pham, C.-M. Luong, M.~Visani, and V.-D. Hoang, ``Deep cnn and data
  augmentation for skin lesion classification,'' in \emph{Asian Conference on
  Intelligent Information and Database Systems}.\hskip 1em plus 0.5em minus
  0.4em\relax Springer, 2018, pp. 573--582.

\bibitem{SANTIN2019}
\BIBentryALTinterwordspacing
M.~Santin, C.~Brama, H.~Théro, E.~Ketheeswaran, I.~El-Karoui, F.~Bidault,
  R.~Gillet, P.~G. Teixeira, and A.~Blum, ``Detecting abnormal thyroid
  cartilages on ct using deep learning,'' \emph{Diagnostic and Interventional
  Imaging}, vol. 100, no.~4, pp. 251 -- 257, 2019. [Online]. Available:
  \url{http://www.sciencedirect.com/science/article/pii/S2211568419300300}
\BIBentrySTDinterwordspacing

\bibitem{CHEVTCHENKO}
\BIBentryALTinterwordspacing
S.~F. Chevtchenko, R.~F. Vale, V.~Macario, and F.~R. Cordeiro, ``A
  convolutional neural network with feature fusion for real-time hand posture
  recognition,'' \emph{Applied Soft Computing}, vol.~73, pp. 748 -- 766, 2018.
  [Online]. Available:
  \url{http://www.sciencedirect.com/science/article/pii/S1568494618305271}
\BIBentrySTDinterwordspacing

\bibitem{srivastava2014dropout}
N.~Srivastava, G.~Hinton, A.~Krizhevsky, I.~Sutskever, and R.~Salakhutdinov,
  ``Dropout: a simple way to prevent neural networks from overfitting,''
  \emph{The Journal of Machine Learning Research}, vol.~15, no.~1, pp.
  1929--1958, 2014.

\end{thebibliography}

\end{document}